\documentclass[12pt]{iopart}

\usepackage{graphicx}
\usepackage[style=numeric,isbn=false,sorting=none]{biblatex}
\bibliography{TaS2-growth-refs.bib}
\usepackage{textcomp,mathcomp,amssymb}
\usepackage{xcolor}
\usepackage{datetime}

\begin{document}

\title{Molecular Beam Epitaxy of 2H-TaS$_2$ few-layers on GaN(0001)}

\author{Constantin Hilbrunner$^1$, Tobias Meyer$^2$, Joerg Malindretos$^1$ and Angela Rizzi$^1$}

\address{$^1$ IV. Physical Institute - Solids and Nanostructures, University of G\"ottingen, D-37077 G\"ottingen, Germany.}
\address{$^2$ Institute of Materials Physics, University of G\"ottingen, D-37077 G\"ottingen, Germany.}
\ead{joerg.malindretos@uni-goettingen.de}

\vspace{10pt}
\begin{indented}
\item[] {(\today , \currenttime)}
\end{indented}

\begin{abstract}
2H-TaS$_2$ few layers have been grown epitaxially onto GaN(0001). A high substrate growth temperature of 825 \textcelsius \ induces best structural properties of the overlayer, as revealed by {\sl in-situ} electron diffraction (RHEED and LEED). The 2D-overlayer grows unstrained right after deposition of a monolayer. However, evidence of pits at the interface is provided by scanning transmission electron microscopy, most probably due to GaN thermal decomposition at the high growth temperature. {\sl In-situ} x-ray photoemission spectroscopy shows core level shifts that are consistently related to electron transfer from the n-GaN(0001) to the 2H-TaS$_2$ epitaxial layer as well as the formation of a high concentration of nitrogen vacancies close to the interface. Further, no chemical reaction at the interface between the substrate and the grown TaS$_2$ overlayer is deduced from XPS, which corroborates the possibility of integration of 2D 2H-TaS$_2$ with an important 3D semiconducting material like GaN.
\end{abstract}

%
% Uncomment for keywords
\vspace{2pc}
\noindent{\it Keywords}: molecular beam epitaxy, tantalum disulfide, gallium nitride, x-ray photoemission spectroscopy
%
% Uncomment for Submitted to journal title message
%\submitto{\JPA}
%
% Uncomment if a separate title page is required
%\maketitle
% 
% For two-column output uncomment the next line and choose [10pt] rather than [12pt] in the \documentclass declaration
%\ioptwocol
%

\submitto{\TDM}
\maketitle

\section{Introduction}
Trigonal prismatic tantalum disulfide (H-TaS$_2$) exhibits in the bulk of its 2H polytype a nearly commensurate $3 \times 3$ charge density wave (CDW) transition below 75~K \cite{Scholz1982,Tidman1974} and a superconducting transition with $T_c = 1$~K \cite{Nagata1992}. As for other layered transition metal dichalcogenides (TMDs), the electronic properties of 2H-TaS$_2$ are dependent on the layer thickness and can be modified by the underlying substrate \cite{DeLaBarrera2018,Yang2018,Fu2020}. For instance, heteroepitaxial growth of a 2H-TaS$_2$ bilayer on a hexagonal-boron nitride (h-BN) substrate produces a robust commensurate CDW order already at room temperature, characterized by a Moir\'e superlattice of $(3 \times 3)$ TaS$_2$ on a $(4 \times 4)$ h-BN unit cell \cite{Fu2020}.

To systematically study the properties of 2H-TaS$_2$, the reproducible fabrication of clean, high-quality layers with controlled thickness is required. Using molecular beam epitaxy (MBE), TMD layers can be synthesized with monolayer control using high-purity elemental sources in an ultrahigh vacuum environment. Successful growth of H-phase TaS$_2$ layers via MBE has been reported on various substrates, including Au(111) \cite{Dombrowski2021} and different van der Waals materials, mostly graphene and h-BN \cite{Shimada1993, Hall2018, Lin2018}.

A major challenge in the growth of single crystalline TMD films is the nucleation of differently orientated islands. Islands rotated to each other coalesce under the formation of grain boundaries, which limit grain size and may modify the electronic properties as compared to the ideal 2D lattice \cite{Batzill:2018, VanDerZande2013, Ma2017}. Due to the in-plane threefold rotational symmetry of single layer TMDs, antiparallel single-crystal triangular island domains (rotation by 60$^\circ$) are energetically degenerate on most surfaces with higher symmetry \cite{Dong2020}. By merging they lead to the formation of structural defects, in particular mirror twin boundaries (MTBs).

Wurtzite GaN offers a matching threefold axis perpendicular to the $c$-plane, and a lattice mismatch of 3.7\%\ to 2H-TaS$_2$. Recently, the first epitaxy of MoS$_2$ on GaN via MBE has been reported; in that case the lattice mismatch is less than 1\%\ \cite{AlKhalfioui2025}. From atomically resolved HR-STEM analyses it is concluded that the 1T phase is more characteristic of the first MoS$_2$ layer which is linked by strong bonds with the GaN substrate, and the 1H phase is generally present between two consecutive MoS$_2$ layers.

In this work, the epitaxial growth of 2H-TaS$_2$ on $c$-plane wz-GaN by MBE and both, their structural and electronic interface properties, have been studied by several {\sl in-situ} techniques. The latter allow a quick feedback between interface properties and the epitaxial growth parameters. An analysis of the XPS core level spectra provides details about electronic modifications within the MBE grown 2H-TaS$_2$ few-layers. Furthermore, scanning transmission electron microscopy (STEM) has been applied to complement the information about structure and chemical composition.

\section{Methods}
2H-TaS$_2$ is grown on commercially available MOCVD grown $c$-plane wz-GaN on sapphire (g-materials) in a Varian Gen II modular MBE system equipped with a reflection high-energy electron diffraction (RHEED) system for growth monitoring and connected to a low-energy electron diffraction (LEED) and an x-ray photoemission spectroscopy (XPS) system for {\sl in-situ} analysis. The substrate wafers were cut into pieces of $10 \mathrm{~mm} \times 10 \mathrm{~mm}$ and their backside was coated with 400~nm of Ti in order to enhance the thermal coupling to the substrate heater and to enable pyrometric temperature measurement. Before being transferred to the reactor, the substrates were heated to 200 \textcelsius \ in the load lock and to 600 \textcelsius \ in a buffer chamber for 10~minutes. The GaN template surface was routinely characterised by RHEED, LEED, XPS, and atomic force microscopy (AFM). 

The 2H-TaS$_2$ layers were grown in a single step process. The substrate was heated to the growth temperature $T_g$, which was chosen between 620 \textcelsius \ and 825 \textcelsius, and was then simultaneously exposed to the Ta and S fluxes. The source materials, Ta (99.99+\%\, MaTeck) and S (99.9995\%\, abcr) were supplied by a small electron beam evaporator (Focus EFM 3s) and a valved cracker cell (Createc V-CRC), respectively. The Ta flux was set to $(2.4 \pm 0.2) \cdot 10^{12}$~atoms cm$^{-2}$ s$^{-1}$. The S flux was characterised by the beam equivalent pressure (BEP) measured with a beam flux monitor ion gauge in substrate position. It was varied between $3 \cdot 10^{-9}$~mbar and  $3 \cdot 10^{-7}$~mbar. The relative S/Ta flux was adjusted to provide stoichiometric composition of the TaS$_2$ epitaxial layers as determined by XPS core level intensity analysis. The growth rate was on the order of 2.5~ML~h$^{-1}$, with 1~ML $\sim 0.6$~nm. Following the deposition, the Ta and S sources were closed simultaneously and the sample was maintained at $T_g$ for 30~minutes before cooling down to room temperature at a rate of $-30$ \textcelsius /min.

XPS measurements were performed using mono\-chromatised Al-K$\alpha$ radiation (1486.3~eV) and a Leybold EA-200 MAX hemispherical analyzer with a pass energy set to 48~eV. The experimental resolution is $\mathrm{FWHM} = 0.7$~eV and the binding energy (BE) scale calibration gives an uncertainty of $\Delta E = 0.15$~eV, according to Ag-3d$_{5/2}$ core level fit and Fermi edge fits on a nickel sample, respectively.

 The photoelectrons are collected along the surface normal from a $4 \mathrm{~mm} \times 7 \mathrm{~mm}$ area with an acceptance angle of $\pm 8^\circ$. For peak fitting, the CasaXPS software package was employed using pseudo-Voigt and asymmetric Voigt like lineshapes along with Shirley backgrounds. For elemental quantification, Scofield relative sensitivity factors for Al-K$\alpha$ radiation are used \cite{Scofield1976}.

TEM samples were prepared in a ThermoFisher Scientific Helios G4 gallium focused ion beam (FIB). To protect the sample surface, a 30~nm thick carbon layer was deposited with the electron beam using an acceleration voltage of 2~kV. Subsequently, further platinum protection bars were deposited with the electron beam (2~kV) and FIB (30~kV). The acceleration voltage during final FIB thinning was set to 2~kV.

Annular dark-field experiments (ADF-STEM) were conducted in an FEI Titan 80-300 E-TEM using an acceleration voltage of 300~kV, a beam current of 42~pA and a convergence semi-angle of 10~mrad. Electron energy loss spectroscopy (EELS) signals were acquired using a Gatan GIF Quantum 965 ER including an Ultrascan 1000XP camera with a dispersion of 1~eV/channel. The corresponding signals were obtained after power law background subtraction.

\section{Results and Discussion}

\begin{figure*}
 \centering
 \includegraphics{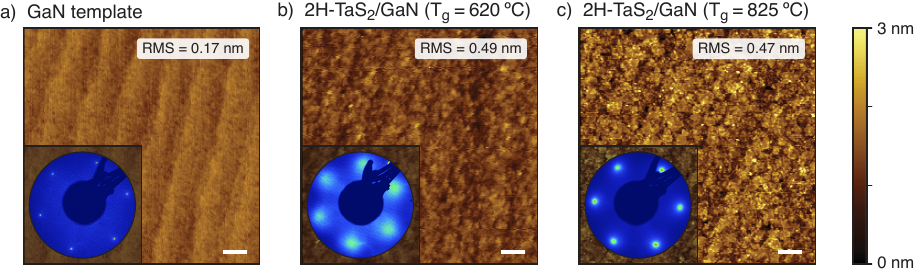}
 \caption{AFM topography measurements (1~\textmu m $\times$ 1~\textmu m) of (a) the GaN substrate, (b) after growth of 2H-TaS$_2$ (6~ML) at $T_g = 620$~\textcelsius \ and (c) after growth at $T_g=825$~\textcelsius. The scale bar length is 100~nm. Corresponding LEED patterns measured at a beam energy of 100~eV are reported in the insets. The slight rotation of the LEED pattern in (c) is due to a rotation of the sample holder.}
  \label{figure:1}
\end{figure*}

Figure~\ref{figure:1} shows AFM topography and LEED patterns of the GaN MOCVD template and two 2H-TaS$_2$ layers grown at different substrate temperatures. The GaN surface in figure~\ref{figure:1}(a) consists of terraces with a width between 70~nm and 120~nm and steps of (0.5 to 0.6)~nm height, consistent with the thickness of a GaN-bilayer ($c=0.5189$~nm in wz-GaN). The corresponding LEED pattern (inset) shows six sharp spots expected for a high quality wurtzite crystal.
The AFM topographies of 6~ML thick 2H-TaS$_2$ layers grown on GaN at $T_g = 620$~\textcelsius \ ({\sl low}) and $T_g = 825$~\textcelsius \ ({\sl high}) are reported in figure~\ref{figure:1}(b) and \ref{figure:1}(c), respectively. In the low temperature topography, a high density of small domains is visible. The AFM measurement only allows to estimate an upper boundary of their size to 20~nm due to the 10~nm curvature radius of the AFM tip. The GaN terraces underneath are still visible. At high growth temperature, the domain size increases to (40 to 70)~nm. 

The LEED diffraction pattern of both 2H-TaS$_2$ layers (insets) consist of six spots, indicating epitaxial growth on GaN(0001). 
The epitaxial relation between the GaN substrate and the grown 2H-TaS$_2$ is discussed below with reference to RHEED experiments. From low to high growth temperature, the LEED spot size clearly decreases. This cannot be explained by the increased domain size alone: its estimation from the LEED spot size yields domains of (7 to 10)~nm, smaller than in the AFM topography. Since for both growth temperatures the diffraction spot FWHM increases linearly with increasing electron beam energy (see supporting information S3), the spot size is attributed to small angle mosaics. These defects form in presence of crystallites connected by small angle (tilt) grain boundaries \cite{VonHoegen1999}.
Furthermore, the intensity from neighboring $\{\bar 1 100\}$ diffraction spots is observed to be slightly modulated according to a three-fold symmetry as revealed in the inset of figure~\ref{figure:1}(c). A pronounced modulation would be expected for diffraction from a monolayer, due to the three-fold symmetry of the single trigonal prismatic cell. On the other hand, diffraction from the multilayer 2H-TaS$_2$ is expected to show a six-fold symmetry. This has indeed been reported for TEM diffraction patterns on single- and multilayer flakes of MoS$_2$, and has been supported by simulations in Ref.~\cite{Brivio:2011}. However, in case of LEED the high surface sensitivity modulates the overall intensity with a prominent contribution from the upmost monolayer, even for multilayer samples.
% NOTE: In our very simple kinematic LEED simulations we do observe a three-fold symmetry for the multilayer structure as well - with and without taking into account the surface sensitivity (the mean free path for 100 eV electrons amounts to ca. 0.5 nm). However, the results depend strongly on the electron acceleration voltage.

\begin{figure}
 \centering
 \includegraphics{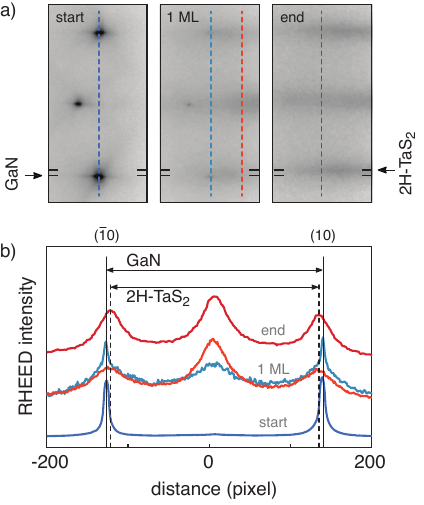}
 \caption{(a) RHEED pattern along GaN [$11\bar20$] at different stages of 2H-TaS$_2$ growth, showing the transition to a streaky pattern of the epitaxial layer. (b) Line profiles along the marked directions in (a) (see text). The intensity profiles have been normalized.}
 \label{figure:2}
\end{figure}

The film growth was monitored online by RHEED. Figure~\ref{figure:2}(a) shows the evolution of the RHEED patterns recorded along the [$11\bar20$] direction of the GaN substrate during high temperature growth. The snapshots are taken at the start of the growth process (left), after the growth of the first TaS$_2$ monolayer (center), and at the end of the growth process of 6~ML (right). The GaN pattern evolves into a streaky 2H-TaS$_2$ pattern after the first monolayer is grown. The diffraction pattern does not change during the growth of further layers, indicating a smooth growing surface. From the RHEED patterns, an epitaxial relation between the GaN substrate and the grown 2H-TaS$_2$ is evident: 2H-TaS$_2$($11\bar20$) $\parallel$ GaN($11\bar20$). The RHEED pattern remains unchanged by scanning the electron beam over the whole sample surface.

In figure~\ref{figure:2}(b), line profiles along the directions indicated in figure~\ref{figure:2}(a) are shown. From the ratio of the distances between the ($\bar10$) and ($10$) reflexes of GaN (deep blue curve; growth start) and 2H-TaS$_2$ (red curve; growth end), the in-plane lattice constant of the overlayer is found to be $a_{\rm 2H-TaS_2} = (0.331 \pm 0.003)$~nm, in agreement with the value of 0.3315~nm expected for bulk 2H-TaS$_2$ \cite{Jellinek1962}. Notably, it does not match the value of 0.336~nm reported for the 1T phase \cite{Jellinek1962}. Furthermore, after the deposition of 1~ML (figure~\ref{figure:2}(a), center panel), in addition to the streaky 2H-TaS$_2$ signal, the GaN substrate pattern is still visible. The analysis of the line profiles intersecting the substrate diffraction peaks (cyan curve) and the 2H-TaS$_2$ signal (orange curve) provides the same value for the in-plane lattice constant. It can therefore be concluded that the first layer grows already fully relaxed, suggesting a van der Waals like growth of 2H-TaS$_2$ on the GaN substrate.

\begin{figure}
 \centering
 \includegraphics{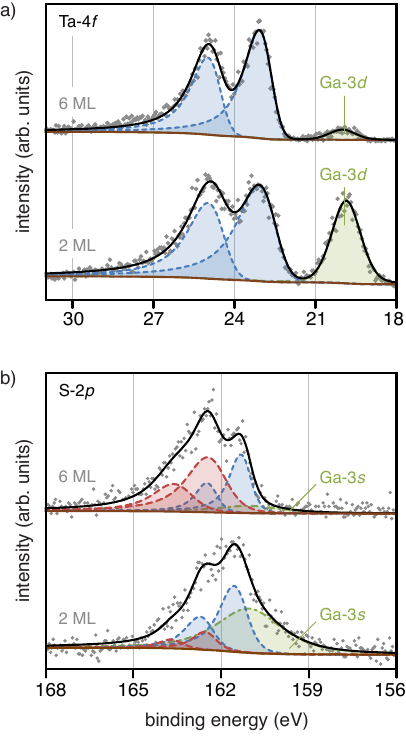}
 \caption{XPS spectra of the Ta-4f (a) and S-2p (b) core levels for 2~ML and 6~ML 2H-TaS$_2$ grown on GaN (dots). Signal from the substrate is detected. The intensity of the spectra is normalized and shifted for clarity. The fitted components are shaded and the black line is the result of the best fit. The S-2p fit required two doublets (blue, short dash and red, long dash). The results of the fit are summarized in table~\ref{tab:1}. Refer to the text for more details.}
 \label{figure:3}
\end{figure}

Core level XPS gives insight into the chemical composition near the interface and within the grown films. The samples were transferred {\sl in-situ} from the MBE reactor to the XPS analysis chamber. No traces of impurities like oxygen or carbon were detected in the recorded spectra. 
In figure~\ref{figure:3}, XPS spectra in the binding energy (BE) region of the Ta-4f and S-2p core levels measured at increasing 2H-TaS$_2$ layer thickness are shown together with best fitted peaks. Due to the thin overlayer thickness of 1.2~nm and 3.6~nm, respectively, photoelectrons from the GaN substrate are clearly detected. The information depth (ID), i.e.\ the sample thickness from which 95\% of the detected signal originates, amounts to (6.5 to 7.5)~nm in the shown binding energy range \cite{Powell:2020}. A summary of the fitted parameters is reported in table~\ref{tab:1}. 

The Ta-4f doublet shows pronounced tails at higher binding energies, therefore it is fitted with an asymmetric line shape. In fact, in the screening process the positive photohole produces excitations in the Fermi sea of conduction electrons. In principle, possible excitation energies range between zero (directly at $E_F$) and the bandwidth of the metal \cite{Huefner:2003}. The band structure of the 2H phase has a characteristic isolated Ta-5d band at the Fermi level, which is partially occupied and which interacts with the Ta-4f photoionization \cite{mattheiss:1973,Eppinga:1976,Lazar:2015}. The Ta-4f$_{7/2}$ binding energy of (22.78 to 22.85)~eV can be compared to the value of 21.74~eV in Ta-metal \cite{XPS-Handbook:1999}; the shift to higher BE is qualitatively expected due to the oxidation of Ta in TaS$_2$, because of the electronegative character of the chalcogen with regard to the metal. The binding energy of 22.85~eV (6~ML) agrees also well within the experimental uncertainty with the 23.0~eV value reported for 2H-TaS$_2$ by Eppinga {\sl et al.}\ \cite{Eppinga:1976}. No other component is needed for a consistent fit of the Ta-4f spectra and therefore we tend to exclude any chemical reaction of Ta at the substrate interface.

The S-2p XPS spectra on the other hand suggest the presence of two doublet components. In fact the measured double peak around 162~eV cannot be reproduced by a single 2p doublet with the due intensity ratio of 2:1. A best fit is obtained with two components, S-2p (blue, short dash) and S$^*$-2p (red, long dash). The latter increases its relative intensity as compared to the main component at increasing TaS$_2$ thickness; its binding energy is about 1~eV higher than for the main component.

\begin{table}[hbt]
  \caption{Fit parameters for the core level components in figure~\ref{figure:3} and figure~\ref{figure:5}. All values are in eV. The FWHM of the single components in a doublet was fitted as a common parameter. Their intensity ratio has been constrained according to the multiplicity of the core-hole final state, which amounts to 4:3 for the Ta-4f and 2:1 for the S-2p. The spin-orbit splitting was fixed at values of $\Delta E_{\rm S-2p}=1.16$~eV and $\Delta E_{\rm Ta-4f}=1.9$~eV \cite{XPS-Handbook:1999}.}
  \vspace{\medskipamount} 
  \resizebox{\columnwidth}{!}
  { %\footnotesize 
  \begin{tabular}{l|cc|cc|ccc}
  	\hline \hline
     Level & \multicolumn{2}{c|}{GaN(0001)} & \multicolumn{2}{c|}{2 ML TaS$_2$} & \multicolumn{3}{c}{$n$ ML TaS$_2$}  \\
     & BE & FWHM & BE & FWHM  & BE & FWHM & $n$\\
    \hline \hline 
    Ta-4f$_{7/2}$ & & & 22.78 & 1.39 & 22.85 &  1.02 & 6\\
    S-2p$_{3/2}$ & & & 161.49 & 1.04 & 161.28 &  0.80 & 6\\
    S$^*$-2p$_{3/2}$ & &  & 162.49 & 0.82 & 162.37 & 1.46 & 6\\
    Ga-3s & & & 161.04 & 2.65 & 161.04 & 2.65 & 6\\
    \hline
    Ga-3d &    19.57   &  1.23    & 19.86 & 1.31 & 20.32 &  1.33 & 4\\
    Ga-2p$_{3/2}$ &     1117.70  &  1.37 &  1117.39    &   1.48     &   1116.94    &   1.23  & 4\\    
     \hline \hline
 \end{tabular}
 } %
 \label{tab:1}
\end{table}

 The S-2p$_{3/2}$ main component at (161.49 to 161.28)~eV is slightly higher than the value of 160.9~eV reported for 2H-TaS$_2$ in Ref.~\cite{Eppinga:1976}. However it should be noted that the broad Ga-3s core level from the substrate lies very close to the S-2p$_{3/2}$ main component, thus impacting on the precision of the peak position. The shifted S$^*$-2p component is likely to result from sulfur in a different coordination than in the bulk. One possibility might be Ta self-intercalation. The 2H-TaS$_2$ stacking sequence implies the presence of octahedrally coordinated interstices in the van der Waals gap. If occupied by Ta atoms, interlayer covalent bonds are created. Generally, S core level shifts in transition metal dichalcogenides are large, while metal ones are absent or small and not easy to detect \cite{Eppinga:1976}. Sulfur atoms at grain boundaries in a configuration with fewer Ta neighbours than in 2D-TaS$_2$ could also contribute to the shifted S$^*$-2p component. A classification of mirror twin boundaries in 2D-TMDs is reported e.g.\ by Batzill in Ref.~\cite{Batzill:2018}. 
 The analysis of the Ta-4f and S-2p intensity ratios, weighted with the sensitivity factors, reproduce the stoichiometry of TaS$_2$ within 10\%, indicating an upper limit to the amount of Ta self intercalation. Therefore the high spectral weight of the S$^*$-2p component in the 6 ML thick layer cannot be entirely assigned to Ta self intercalation.

\begin{figure}
 \centering
 \includegraphics{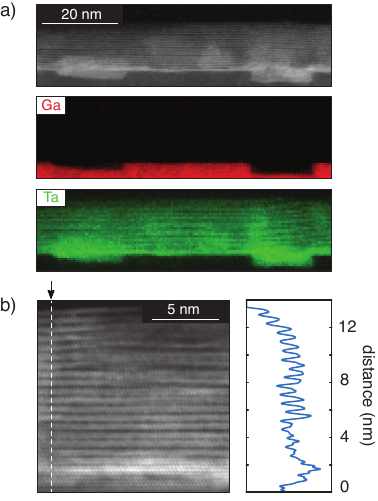}
 \caption{(a) Cross section ADF-STEM image of an 18~ML thick 2H-TaS$_2$ layer grown on GaN. Below, EELS signals from the Ga L-edge and Ta M-edge after background subtraction are shown. (b) Higher magnification image with an intensity profile along the marked direction on the right.}
 \label{figure:4}
\end{figure}

In figure~\ref{figure:4}(a), a cross section STEM image and EELS maps of 18~ML 2H-TaS$_2$ grown on GaN are reported. The 2H-TaS$_2$ grows in a layered structure. In the GaN substrate, pits of ca. 15~nm length and (1 to 2.5)~nm depth are observable. They are attributed to GaN thermal decomposition due to the elevated substrate temperature during growth. It must be noted that the overlayer growth process onto GaN occurs (i) in absence of a nitrogen atmosphere, as opposed to the case of standard GaN MBE growth where the nitrogen background pressure is on the order of $10^{-5}$~mbar, and (ii) with a low growth rate of about 2.5~ML/h. Under these conditions a significant evaporation of nitrogen from the sample surface may certainly be an issue.  The EELS measurements show a clear separation of Ga and Ta without intermixing. The intense Ta EELS signal in the pits and at the direct 2H-TaS$_2$/GaN interface indicate the formation of a Ta-rich phase.  A higher magnification cross-sectional STEM image is shown in figure~\ref{figure:4}(b). The slightly blurred Ta intensity compared to the sharp Ga signal in the substrate suggests disorder in the grown 2H-TaS$_2$ layers. From the intensity profile on the right, the 2H-TaS$_2$ monolayer spacing is extracted to be ($0.63\pm 0.04$)~nm, which compares well with literature values of 0.60~nm ($= 0.31\rm{~nm} + 0.29\rm{~nm}$), corresponding to the TaS$_2$ sandwich plus the height of the van der Waals gap, respectively \cite{Meetsma:1990}. For some area on the right, the intensity profile inside the van der Waals gap is smeared out. Beyond structural defects, self-intercalation of Ta, which  has been observed under slightly Ta-rich growth conditions in MBE \cite{Zhao2020} and CVD \cite{Han2024}, might explain this observation. 

Low temperature LEED measurements were performed to check for indications of a charge density wave phase, which emerges below 75~K in bulk 2H-TaS$_2$.  It is worth noting that no LEED evidence of the CDW transition in 2H-TaS$_2$ has been reported in the literature so far. According to the small atomic displacement \cite{Zhao2018}, the intensity of the superstructure due to the periodic lattice distortion in 2H-TaS$_2$ is expected to be reduced by a factor of 25 as compared with 1T-TaS$_2$. The access to a low-temperature, spatially resolved LEED setup with a microchannel plate detector provided best experimental conditions with high surface sensitivity \cite{Kurtz:2024} for the aim of revealing the occurrence of the expected superstructure \cite{Scholz1982, Hall2019}.

\begin{figure}
 \centering
 \includegraphics[scale=0.9]{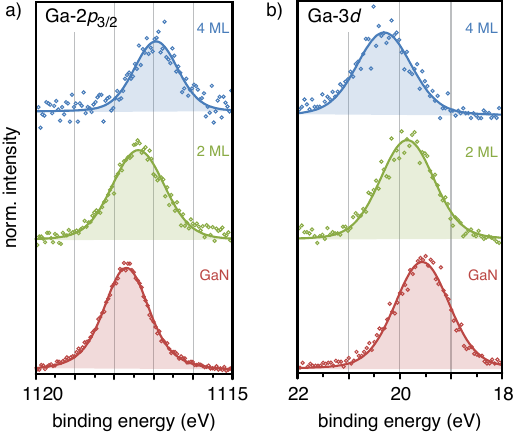}
 \caption{Evolution of XPS spectra measured {\sl in-situ} on the bare GaN substrate and after deposition of 2H-TaS$_2$. (a) Ga-2p$_{3/2}$ and (b) Ga-3d. The fitted parameters are reported in table~\ref{tab:1}. The background corrected intensity of the spectra is normalized and shifted for clarity.}
 \label{figure:5}
\end{figure}

However, no LEED signal corresponding to a CDW induced superstructure was detected down to a temperature of 30~K (see supporting information S4). It is worth noting that, in contrast to their semiconducting counterparts \cite{Gao2016}, metallic TMDs are prone to oxidation under ambient conditions or during synthesis \cite{Zhao2018, Pathan2021}. However, the transfer to the low-T LEED apparatus with exposure to air for about 5~min did not cause any significant oxidation, as checked with XPS. Therefore we exclude 2H-TaS$_2$ oxidation to hamper the CDW transition. While structural properties as strain, defects and Ta self-intercalation cannot be excluded as a cause for the absence of a CDW phase, it is known that charge doping can suppress the CDW in 2H-TaS$_2$ \cite{Hall2019}. Therefore a charge transfer between the unintentionally doped n-GaN template and the MBE grown 2H-TaS$_2$ has been investigated by XPS. Spectral features arising from the GaN semiconducting substrate are measured {\sl in-situ} after deposition of different TaS$_2$ thicknesses and are shown in figure~\ref{figure:5}.
% NOTE: It may be worth mentioning, that it is not clear in principle whether the sensitivity of our LEED would allow the observation of the CDW superstructure in 2H-TaS2, since the amplitude is expected to be much smaller than in 1T-TaS2 [PhD thesis by CH].

From figure~\ref{figure:5}(a) it is evident that the Ga-2p$_{3/2}$ core level shifts to lower binding energies at increasing overlayer thickness, as compared to the bare substrate. A shift in the opposite direction --to higher binding energy-- is measured for the Ga-3d spectra as revealed in figure~\ref{figure:5}(b). Electron transfer from the n-GaN to the 2H-TaS$_2$ is expected at the interface based on the relative values of the GaN and 2H-TaS$_2$ work functions with values of 4~eV \cite{Kampen:1998} and 5.2~eV \cite{Shimada1993}, respectively. The relative shift of $-0.76$~eV of the Ga-2p$_{3/2}$ core level to lower BE for the 4~ML thick overlayer is indeed consistent with an upwards band bending in GaN at the metal/semiconductor interface (Schottky contact). Note that the information depth for the Ga-2p$_{3/2}$ photoelectrons amounts to 2.8~nm and therefore the signal comes from the very interface. The Ta-4f$_{7/2}$ core level remains constant, within the experimental uncertainty (table~\ref{tab:1}), due to the effective screening of the interface dipole on the side of the metal.

Obviously, the Ga-3d shift of $+0.75$~eV (4~ML) is in the opposite direction and cannot be explained only in terms of band bending. It has been discussed that the proper treatment of the cation 3d states as a band is important in GaN \cite{Fiorentini:1993}. In fact, the free-atom energy eigenvalue $\epsilon_{3d}^{\mathrm{Ga}} = -20$~eV is close to resonance with $\epsilon_{2s}^{\mathrm{N}}= -18.37$~eV. As a consequence the center of mass of the hybridized $sd$ crystal levels is raised in energy with respect to the center of mass of the corresponding atomic level, probably due to an enhanced closed-shell repulsion of the hybridized core eigenstates \cite{Fiorentini:1993}. The observed thermal decomposition of GaN after deposition of 2H-TaS$_2$ as revealed by the STEM-EELS analysis in figure~\ref{figure:4} is caused by evaporation of nitrogen from the surface. Concentrations of nitrogen vacancies above $10^{-19}$~cm$^{-3}$ and enhanced self-diffusion onto the nitrogen sub-lattice have been estimated by {\sl ab initio} calculations for negatively charged nitrogen vacancies V$_{\mathrm{N}}$ in n-GaN at high temperature (900 \textcelsius) \cite{Ganchenkova:2006}. As 2H-TaS$_2$ MBE growth proceeds, the missing N neighbors and corresponding $sd$ hybridisation reduce the closed-shell repulsion causing the Ga-3d energy band to shift to higher binding energies. An energy band shift to lower energy in GaN:V$_{\mathrm{N}}$ has also been calculated by density functional theory \cite{Cai:2017}. By subtracting the band bending shift of $-0.76$~eV, as determined from the Ga-2p$_{3/2}$ core level, from the measured Ga-3d shift of $+0.75$~eV, an overall Ga-3d band shift of $\Delta_{\rm BE} = +1.51$~eV is deduced. This is qualitatively in agreement with the band structure calculations of GaN:V$_{\mathrm{N}}$ \cite{Cai:2017}. Therefore, we conclude that the shift of the deeper Ga-2p core levels only reflects the band bending effect, whereas the Ga-3d electron states, which cannot be treated as "frozen" core levels in GaN, are also affected by energy band shifts.

\section{Summary}
Few-layer 2H-TaS$_2$ is grown at a high substrate temperature ($T_g=825$ \textcelsius) on GaN by molecular beam epitaxy. RHEED and LEED experiments show the van der Waals epitaxy of (unstrained) 2H-TaS$_2$ onto GaN(0001), already from the first monolayer. Because of the high growth temperature, pits of thermally decomposed GaN substrate are observed in cross section STEM measurements. {\sl In-situ} core level XPS allows a detailed study of the electronic interface properties. While no reacted component of Ta and Ga is seen in XPS spectra, a second sulfur 2p doublet (S$^*$-2p) is revealed, which is assigned to sulfur bonding either to self-intercalated Ta or at grain boundaries. Further, the observed shift of the Ga-2p$_{3/2}$ core level reflects an upwards band bending in GaN near the interface. Hence, electron transfer from the GaN to the 2H-TaS$_2$ occurs. Whether the 2H-TaS$_2$ electron doping is the reason for the absence of the charge density wave transition expected below $75$~K remains to be clarified, e.g.\ by MBE growth on either semi-insulating or p-GaN. In conclusion, these results demonstrate that GaN is a promising substrate for the MBE growth of highly oriented wafer-scale TMD layers. Further optimization of the MBE growth to avoid GaN decomposition at the interface, for instance by TaS$_2$ growth under N$_2$ atmosphere, is required.

\section*{Acknowledgements}
The authors are grateful to Claus Ropers (Max Planck Institute for Multidisciplinary Sciences and University of G\"ottingen) for providing the tantalum and sulfur MBE sources, to J\"urgen Bl\"asing (Otto von Guericke University Magdeburg) for HR-XRD experiments and to Alp Akbiyik for the low-temperature LEED experiments. The use of equipment in the “Collaborative Laboratory and User Facility for Electron Microscopy” (CLUE -- www.clue.physik.uni-goettingen.de) is gratefully acknowledged. Thomas Lehmann is acknowledged for technical assistance.

\section*{References}
\printbibliography[heading=none]

% SUPPLEMENTARY FIGURES
\clearpage

\renewcommand{\thesection}{S}
\renewcommand\thefigure{\thesection \arabic{figure}} 
\setcounter{figure}{0}    

\title{Supporting Information}

%\section*{Supporting Information}
\begin{figure*}[htp]
 \centering
 \includegraphics{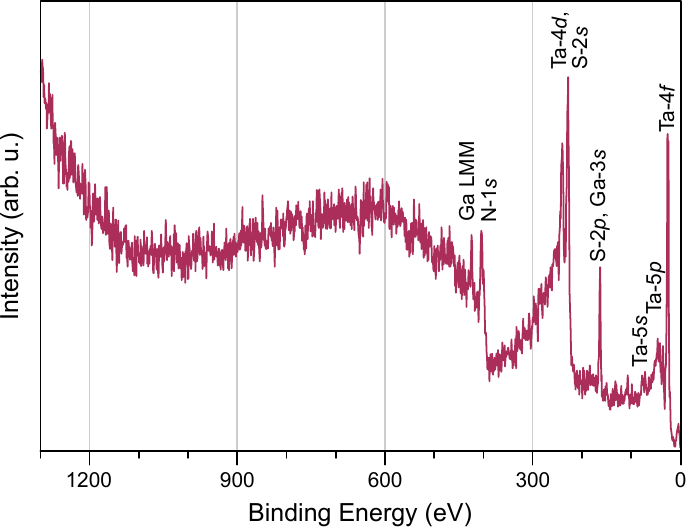}
 \caption{XPS survey spectrum of 6~ML 2H-TaS$_2$ grown on GaN measured with a pass energy of $192$~eV.}
 \label{figure:s1}
\end{figure*}
\begin{figure*}[htp]
 \centering
 \includegraphics{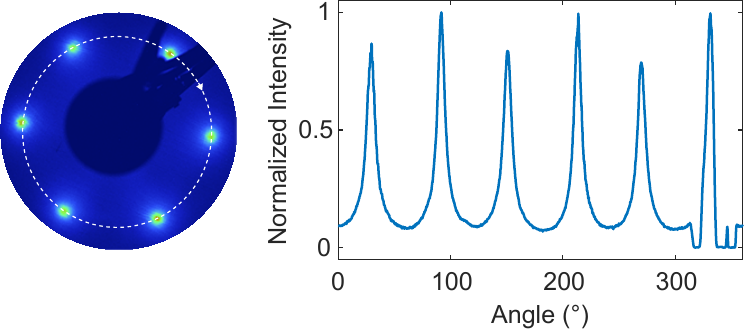}
 \caption{LEED pattern of 6~ML 2H-TaS$_2$ layers grown on GaN acquired with a beam energy of 100~eV (left) with the angular intensity profile extracted along the dashed white line (right).}
 \label{figure:s2}
\end{figure*}
\begin{figure*}[htp]
 \centering
 \includegraphics{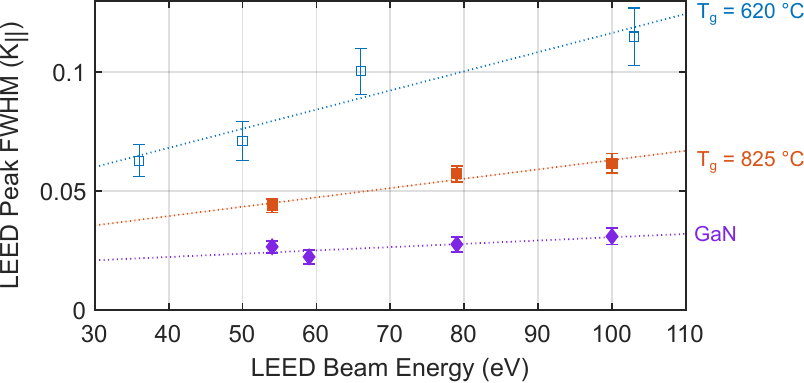}
 \caption{Increase of the relative LEED spot size with increasing beam energy for 6~ML 2H-TaS$_2$ layers grown on GaN at $T_g = 620$~\textcelsius \ and $T_g = 825$~\textcelsius. For comparison, a measurement on the GaN template is shown.}
 \label{figure:s3}
\end{figure*}
\begin{figure*}[htp]
 \centering
 \includegraphics{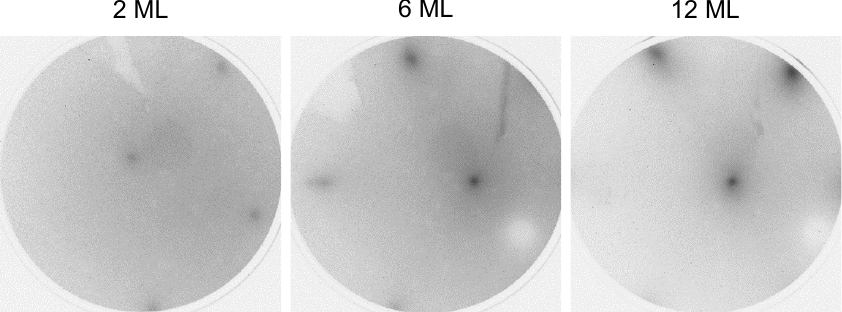}
 \caption{Low temperature LEED patterns of MBE grown 2H-TaS$_2$ layers with different thicknesses measured at $T = 30$~K at a beam energy of $85$~eV. The images are centered around the zeroth-order diffraction spot. No CDW-related sattelite spots around the main reflexes are observable.}
 \label{figure:s4}
\end{figure*}

\end{document}